\documentclass[12pt,eadjoint tfnpsf]{article}

\usepackage{amsmath,amsbsy,amssymb,latexsym}
\usepackage{graphicx}

\newcommand{\BE}{\begin{equation}}
\newcommand{\EE}{\end{equation}}
\newcommand{\BEA}{\begin{eqnarray}}
\newcommand{\EEA}{\end{eqnarray}}
\def\12{\frac{1}{2}}
\def\bea{\begin{eqnarray}}
\def\eea{\end{eqnarray}}
\def\ba{\begin{array}}
\def\ea{\end{array}}
\def\be{\begin{equation}}
\def\ee{\end{equation}}

\def\one-loop{\mbox{\scriptsize one-loop}}

\def\G{\Gamma}

\catcode`\@=11

\@addtoreset{equation}{section}
\def\theequation{\arabic{section}.\arabic{equation}}

\def\@normalsize{\@setsize\normalsize{15pt}\xiipt\@xiipt
\abovedisplayskip 14pt plus3pt minus3pt%
\belowdisplayskip \abovedisplayskip
\abovedisplayshortskip \z@ plus3pt%
\belowdisplayshortskip 7pt plus3.5pt minus0pt}

\def\small{\@setsize\small{13.6pt}\xipt\@xipt
\abovedisplayskip 13pt plus3pt minus3pt%
\belowdisplayskip \abovedisplayskip
\abovedisplayshortskip \z@ plus3pt%
\belowdisplayshortskip 7pt plus3.5pt minus0pt
\def\@listi{\parsep 4.5pt plus 2pt minus 1pt
\itemsep \parsep
\topsep 9pt plus 3pt minus 3pt}}

\def\underline#1{\relax\ifmmode\@@underline#1\else
$\@@underline{\hbox{#1}}$\relax\fi}
\@twosidetrue
\relax

\catcode`@=12

\catcode`\@=11

\def\section{\@startsection{section}{1}{\z@}{3.5ex plus 1ex minus
.2ex}{2.3ex plus .2ex}{\large\bf}}

\def\thesection{\Roman{section}.}

\def\appendix{\setcounter{section}{0}
\def\thesection{Appendix }
\def\theequation{\Alph{section}.\arabic{equation}}}

\catcode`\@=12

\relax

\def\figcap{\section*{Figure Captions\markboth
{FIGURECAPTIONS}{FIGURECAPTIONS}}\list
{Fig. \arabic{enumi}:\hfill}{\settowidth\labelwidth{Fig. 999:}
\leftmargin\labelwidth
\advance\leftmargin\labelsep\usecounter{enumi}}}
 \relax
\def\tablecap{\section*{Table Captions\markboth
{TABLECAPTIONS}{TABLECAPTIONS}}\list
{Table \arabic{enumi}:\hfill}{\settowidth\labelwidth{Table 999:}
\leftmargin\labelwidth
\advance\leftmargin\labelsep\usecounter{enumi}}}
 \relax
\def\reflist{\section*{References\markboth
{REFLIST}{REFLIST}}\list
{[\arabic{enumi}]\hfill}{\settowidth\labelwidth{[999]}
\leftmargin\labelwidth
\advance\leftmargin\labelsep\usecounter{enumi}}}
 \relax

\newskip\humongous \humongous=0pt plus 1000pt minus 1000pt

\newif\ifdtup

\def\Tr{\mathop{\rm Tr}}

\def\beq{\begin{equation}}
\def\eeq{\end{equation}}

\def\beqn{\begin{eqnarray}}
\def\eeqn{\end{eqnarray}}
\relax

\def\G2{{\; \rm GeV/}c2}
\def\G{\; \rm GeV}

\def\dotx{\dotx{\dot\overline{x}}}

\relax

\newcommand\CD{{\mathcal D}}
\newcommand\CF{{\mathcal F}}

\newcommand\CN{{\mathcal N}}

\newcommand\CW{{\mathcal W}}

\def\ee{{\rm e}^}

\renewcommand{\thefootnote}{\fnsymbol{footnote}}
\begin{document}
%%%%%%%%%%%%%%%%%%%%%%%%%%%%%%%%%%%%%%
%
% title page
%
%%%%%%%%%%%%%%%%%%%%%%%%%%%%%%%%%%%%%%
\begin{titlepage}

\begin{flushright}
\normalsize
%\filename
~~~~
April, 2007 \\
OCU-PHYS 262 \\
\end{flushright}

%\vfill
%
\begin{center}
{\Large\bf Deformation of Dijkgraaf-Vafa Relation via \\
  Spontaneously Broken ${\mathcal N}=2$ Supersymmetry }
\end{center}

\vfill

\begin{center}
{%
H. Itoyama$^{a,b}$\footnote{e-mail: itoyama@sci.osaka-cu.ac.jp}
\quad and \quad
K. Maruyoshi$^a$\footnote{e-mail: maruchan@sci.osaka-cu.ac.jp}
}
\end{center}

\vfill

\begin{center}
$^a$ \it Department of Mathematics and Physics,
Graduate School of Science\\
Osaka City University\\
\medskip

$^b$ \it Osaka City University Advanced Mathematical Institute
(OCAMI)

\bigskip

3-3-138, Sugimoto, Sumiyoshi-ku, Osaka, 558-8585, Japan \\

\end{center}

\vfill

%%%%%%%% abstract %%%%%%%%%%%%%%%%%%%%%
\begin{abstract}
  It is known that the fermionic shift symmetry of the ${\mathcal N} =1$, 
  $U(N)$ gauge model with a superpotential of an adjoint chiral superfield
  is replaced by the second (spontaneously broken) supersymmetry 
  in the ${\mathcal N} =2$, $U(N)$ gauge model with a prepotential and Fayet-Iliopoulos parameters.
  Based on a diagrammatic analysis, we demonstrate how the well-known form of the effective superpotential 
  in the former model is modified in the latter. 
  A set of two equations on the one-point functions stating the Konishi anomaly is modified accordingly.

\end{abstract}

\vfill

\setcounter{footnote}{0}
\renewcommand{\thefootnote}{\arabic{footnote}}

\end{titlepage}

%%%%%%%%%%%%%%%%%%%%%%%%%%%%%%%%%%%%%%%%%%%%%%%%%%%%%%%%%%%%%%%%%%%%%%
\section{Introduction}
  For more than two decades, effective superpotential has been a central object 
  in the nonperturbative study of ${\mathcal N}=1$ supersymmetric theories. 
  This object is protected from perturbative corrections in the conventional sense \cite{GSR},  
  and yet receives important nonperturbative corrections (see for example \cite{VY, ADS}). 
  In recent years, analyses from superstring theory have revealed 
  an interesting perturbative window into nonperturbative physics
  with the use of the gluino condensate superfield variable \cite{Vafa, CIV, CV, DV}.
  In \cite{CDSW}, field theoretic discussion
  based on the model with $U(N)$ gauge group 
  and rigid ${\mathcal N}=1$ supersymmetry (see eq. (\ref{S1}) for its action $S_{\CN=1}$)
  is given and this is in accord with the string theory based developments.
  
  Superstring theory, on the other hand, insists upon maximally extended
  supersymmetry with no adjustable parameter. 
  A scenario that one may draw is 
  that this extended supersymmetry becomes spontaneously broken to ${\mathcal N}=1$. 
  Along this vein, a field theory model 
  with $U(N)$ gauge group and rigid ${\mathcal N}=2$ supersymmetry spontaneously broken to ${\mathcal N}=1$
  has been introduced in \cite{FIS1, FIS2, FIS3} (see eq. (\ref{S2}) for its action $S_{\CN=2}$), 
  generalizing the abelian counterpart of \cite{APT}. 
  (See also \cite{sugra} for $\CN=2$ supergravity and \cite{KMG} for related discussions.)
  Several properties of this model have been derived.
  
  In this letter, we make a first analysis on the interplay between
  the effective superpotential and partially as well as spontaneously broken
  ${\mathcal N}=2$ supersymmetry, 
  shedding a light upon the comparison of the two models mentioned above.
  A key aspect of this comparison is that the fermionic shift symmetry of $S_{{\mathcal N}=1}$ 
  gets replaced by the second (spontaneously broken) supersymmetry of $S_{{\mathcal N}=2}$. 
  In fact, this is one of the original motivations/results of \cite{FIS1}.
  
  The fermionic shift symmetry of $S_{{\mathcal N}=1}$ supplies 
  the well-known formula \cite{DV, CDSW} constraining the form of the effective superpotential, 
  which is originally proposed from flux compactification of string theory \cite{GVW, TV}.
  Based on a diagrammatic analysis \cite{DGLVZ} (for a review see \cite{AFH}), 
  we are able to state
  how this form undergoes modifications in the model $S_{{\mathcal N}=2}$.
  After giving a few accounts of the model in the next section, 
  we present a diagrammatic analysis of $W_{eff}$ in section III. 
  Our final understanding is summarized in eq. (\ref{Weffh-1}). 
  This is followed by a computation of the two-loop contribution to $W_{eff}$ in section IV.
  In the final section, we derive a set of two equations on
  the two generating functions $R(z)$ and $T(z)$ of the one-point functions, 
  generalizing the argument based on the chiral ring and the Konishi anomaly in \cite{CDSW}. 
  We observe a modification from that given in \cite{CDSW} here as well.

%%%%%%%%%%%%%%%%%%%%%%%%%%%%%%%%%%%%%%%%%%%%%%%%%%%%%%%%%%%%%%%%%%%%%%%%
\section{The $U(N)$ gauged model with spontaneously broken $\CN=2$ supersymmetry}
  Let us briefly recall a few ingredients of the model, 
  which are needed in what follows.
  The action \cite{FIS1} given in the Wess-Zumino gauge can be written as 
    \bea
    S_{\CN=2}   
    &=&    \int d^4 x d^4 \theta 
           \left[
         - \frac{i}{2} {\rm Tr} 
           \left(  \bar{\Phi} e^{ad V} 
           \frac{\partial \CF(\Phi)}{\partial \Phi}
         - h.c.
           \right)
         + \xi V^0 
           \right] 
           \nonumber \\
    & &  + \left[
           \int d^4 x d^2 \theta
           \left(
         - \frac{i}{4} 
           \frac{\partial^2 \CF(\Phi)}{\partial \Phi^a \partial \Phi^b}
           \CW^a \CW^b
         + e \Phi^0
         + m \frac{\partial \CF(\Phi)}{\partial \Phi^0}
           \right)
         + h.c.
           \right],      
           \label{S2}
    \eea
  where $V= V^a t_{a}$ and  $\mathcal{W}^\alpha$
  are the vector superfield and the gauge superfield strength respectively and
  $\Phi = \Phi^a t_a$ ($a=0, 1, \dots, N^2-1)$ is the chiral superfield
    \footnote{$a=0$ corresponds to the overall $U(1)$ part.}. 
  There are three Fayet-Iliopoulos parameters $(e, m, \xi)$ which are all real.
  For simplicity, we choose the prepotential as a single trace function of degree $n+2$:
  $ \CF(\Phi) = \sum_{k=1}^{n+1} g_k {\rm Tr} \Phi^{k+1} / (k+1)!$.
  While this action is shown to be invariant under the $\CN=2$ supersymmetry
  transformations \cite{FIS1, FIS2},  the vacuum breaks half of 
  the $\CN=2$ supersymmetries.
  Extremizing the scalar potential, we obtain the condition 
    $\langle \frac{\partial^2 \CF}{\partial \Phi^0 \partial \Phi^0} \rangle = 
     - ( e \pm i \xi )/m $,
  which is a polynomial of order $n$ and this determines the expectation value of the scalar field.
 
  The action $S_{\CN=2}$ in (\ref{S2}) is to be compared with
  that of the $\CN=1$, $U(N)$ gauge model with a single trace tree level superpotential $W(\Phi)$:
    \bea 
    S_{\CN=1}
     =     \int d^4 x d^4 \theta 
           \Tr \bar{\Phi} e^{ad V} \Phi
         + \left[
           \int d^4 x d^2 \theta
           \Tr \left(
           i \tau \CW \CW
         + W(\Phi)
           \right)
         + h.c. 
           \right],
           \label{S1}
    \eea
  where $\tau$ is a complex gauge coupling $\tau = \theta/2 \pi + 4 \pi i / g^2$.
  
  In \cite{FIS1}, it is checked that the second supersymmetry reduces 
  to the fermionic shift symmetry in the limit $m \rightarrow \infty$.
  The action $S_{\CN=2}$ in fact reduces to $S_{\CN=1}$
  in the limit $m, e, \xi \rightarrow \infty$ with $m g_k$ ($k \geq 2$) fixed \cite{Fujiwara}.
  We show that our result reduces to that of \cite{DV, DGLVZ} in this limit. 

%%%%%%%%%%%%%%%%%%%%%%%%%%%%%%%%%%%%%%%%%%%%%%%%%%%%%%%%%%%%%%%%%%%%%%%%%%%%
\section{Diagrammatic analysis of the effective superpotential}
  In this letter, we consider the matter-induced part of 
  the effective superpotential by integrating out the massive degrees of freedom $\Phi$:
    \bea
    e^{i \int d^4 x (d^2 \theta W_{eff} + h.c. + d^4 \theta ({\rm nonchiral~terms}))}
     =     \int \CD \Phi \CD \bar{\Phi} e^{i S_{\CN=2}}.
           \label{Weff}
    \eea                                                          
  Let us take $\mathcal{W}^\alpha$ (or $V$) as the background field 
    \footnote{The simplest background is that consisting of 
    a vanishing gauge field $A_\mu$ and a constant gaugino $\lambda^\alpha$, 
    which satisfies $\{ \lambda^\alpha, \lambda^\beta \} = 0$ \cite{AFH}.
    This configuration implies that traces of more than two $\CW$ vanish.}.
  We consider the case of unbroken $U(N)$ gauge group.
  For simplicity,  we choose $\left< \Phi \right> = 0$ by setting
  $g_1 = - (e \pm i \xi)/m$.
  
  We are interested in the holomorphic superpotential which does not contain 
  the anti-holomorphic couplings $\bar{g}_k$.
  We can take $\bar{g}_k=0$ for $k\geq3$ without loss of generality.
  Collecting the $\bar{\Phi}$ dependent terms,  we obtain
    \bea
    S_{\bar{\Phi}}
    &=&    \int d^4 x d^4 \theta 
           \frac{- i}{2} {\rm Tr} 
           \left[ \bar{\Phi} e^{ad V} 
           \frac{\partial \CF(\Phi)}{\partial \Phi}
         - (\bar{g}_1 \bar{\Phi} + \frac{\bar{g}_2}{2} \bar{\Phi}^2) e^{ad V} \Phi 
           \right] 
         + \int d^4 x d^2 \bar{\theta} \frac{m \bar{g}_2}{2} {\rm Tr} \bar{\Phi}^2
           \nonumber \\
    &=&    \int d^4 x d^4 \theta 
           {\rm Tr}
           \left[
           \tilde{\Phi} \bar{g}_2 \left( - \frac{2 m}{\nabla^2} + \frac{i}{4} \Phi \right) \tilde{\Phi}
         + \frac{i}{2} 
           \left(
           \bar{g}_1 \Phi - \frac{\partial \CF}{\partial \Phi}
           \right) \tilde{\Phi}
           \right].
           \label{Sbar}
    \eea
  In the last expression, we have introduced a covariantly anti-chiral superfield $\tilde{\Phi} = \bar{\Phi} e^{ad V}$, 
  which satisfies $\nabla_\alpha \tilde{\Phi} = 0$ ($\nabla_\alpha = e^{- ad V} D_\alpha e^{ad V}$).
  Eq. (\ref{Sbar}) is quadratic in $\tilde{\Phi}$ and can be integrated straightforwardly.
  As a result, we obtain the following terms, 
    \bea
    \frac{1}{16 \bar{g}_2}
    \left(
    \bar{g}_1 \Phi - \frac{\partial \CF}{\partial \Phi}
    \right)
    \left(
    - \frac{2 m}{\nabla^2} + \frac{i}{4} \Phi
    \right)^{-1}
    \left(
    \bar{g}_1 \Phi - \frac{\partial \CF}{\partial \Phi}
    \right)
     =     \frac{({\rm Im} g_1)^2}{8 m \bar{g_2}} \Phi \nabla^2 \Phi
         + \dots,
    \eea
  where $\ldots$ denotes the higher order interaction terms, which we will not consider here.
  Indeed, these interaction vertices are higher order in $m^{-1}$ 
  compared to the vertices which we consider below.
  These contribute to our main result (\ref{Weffh-1}) 
  as higher order corrections in $m^{-1}$ and do not spoil our conclusion 
  that the effective superpotential is modified from the case of $S_{\CN=1}$ (\ref{S1}).

  Replacing $d^2 \bar{\theta}$ integration by  $- \bar{\nabla}^2/4$ 
  and collecting the terms which are not in $S_{\bar{\Phi}}$, 
  we obtain an action after the $\bar{\Phi}$ integration
    \footnote{In eq. (\ref{SPhi}), 
              it is understood that the generating functional has a renormalized perturbation expansion
              in which a nonvanishing tadpole is always canceled by a nonvanishing value of the source coupled to $\Phi$.
              This implies that the tadpole can in practice be ignored.}:
    \bea
    \int d^4 x d^2 \theta {\rm Tr}
    \left[ - \frac{({\rm Im} g_1)^2}{32 m \bar{g}_2} \Phi \bar{\nabla}^2
    \nabla^2 \Phi + m \sum_{k=2}^{n+1} \frac{g_k}{k!} \Phi^k  - \frac{i}{4}
    \sum_{k=3}^{n+1} \sum_{s=0}^{k-1} \frac{g_k}{k!} (\CW \Phi^s \CW \Phi^{k-1-s})
    \right].
    \label{SPhi}
    \eea
  The first two terms are already present in the integrations with regard to
  the action $S_{\CN=1}$ (\ref{S1}).
  The last term is new and originates from the gauge kinetic term in eq. (\ref{S2}).
  As we will see below, this last term does contribute to
  the effective superpotential and  becomes responsible for the violation
  of the well-known relation \cite{DV, CDSW} between the effective superpotential of
  the gauge theory and  the planar free energy of the matrix model having 
  the tree level (bare) superpotential as its potential. 

  After rescaling $\Phi \rightarrow a\Phi$ with $a^2=m \bar{g}_2/({\rm Im} g_1)^2$, 
  the quadratic part of the action (\ref{SPhi}) reduces to
    \bea
    \frac{1}{2} \Phi
    \left( 
    - \square + m' + \frac{1}{2} ad\CW^\alpha D_\alpha 
    \right) \Phi
    - \frac{i g'_3}{2} (2 \CW \CW \Phi^2 + \CW \Phi \CW \Phi),
    \nonumber
    \eea
  where we have used the relation
  $\bar{\nabla}^2 \nabla^2 \Phi = 16( \square \Phi - ad\CW^\alpha D_\alpha \Phi/2)$
  and introduced  $ m'= a^2 m g_2$ and $g'_3 = a^2 g_3/12$.
  The propagator in the momentum space is
    \bea
    \Delta(p, \pi)
     =     \int_0^{\infty} d s 
           e^{-s (p^2 + m' + \frac{1}{2} ad\CW^\alpha \pi_\alpha - i g'_3 M}).
           \nonumber
    \eea
  The Grassmann momentum $\pi^\alpha $ is Fourier transformation of
  superspace coordinate $\theta^\alpha$ and  the matrix $M$ is
    \bea
    M_{abcd}
     =     (\CW \CW)_{da} \delta_{bc} + (\CW \CW)_{bc} \delta_{da} + \CW_{da} \CW_{bc},
           \label{M}
    \eea
  where we have exhibited the gauge index dependence explicitly.
  This matrix is not present in the propagator of \cite{DGLVZ}.
  Using eq. (\ref{M}), we are able to insert $\CW$ without involving the momentum $\pi^\alpha$.
  
  The interaction terms in eq. (\ref{SPhi}) are divided into the following two types:
    \bea
    {\rm type~I.}
    & &    ~~~~~~~
           m \frac{g_k a^k}{k!} \Tr \Phi^k, 
           ~~~~~~~~~~
           k = 3, \ldots, n+1.
           \nonumber \\
    {\rm type~II.}
    & &    ~~~~
         - \frac{i}{4} \sum_{s=0}^{k-1}
           \frac{g_k a^{k-1}}{k!} \Tr (\CW \Phi^s \CW \Phi^{k-1-s}),
           ~~~~~~~~~~
           k = 4, \ldots, n+1.
           \nonumber
    \eea
  Type I vertices are already present in \cite{DGLVZ}.
  Type II vertices are not present in \cite{DGLVZ}.
  They insert two $\CW$ in specific ways.

  Before going on to consider loop diagrams, 
  let us first demonstrate that we have only to consider planar diagrams 
  in our case as well \cite{DGLVZ, AFH}.
  For a given diagram,  we denote by $V$ the number of vertices,
  by $P$ the number of propagators and by $h$ the number of holes (or index loops).
  There are $V$ sets of chiral superspace integrations from $V$ vertices.
  One of them becomes the chiral superspace integration over the effective superpotential,  
  and the number of remaining $\pi^\alpha$ momentum integrations is $P - V + 1$.
  These Grassmann integrations must be saturated by $\frac{1}{2} ad \CW^\alpha \pi_\alpha$ terms in the propagators.
  Furthermore, we can freely insert $\CW$ 
  both from the $M$ terms in the propagators and from the type II vertices.
  If we denote the number of these additional insertions by $2 \alpha$, 
  the total number of $\CW$ insertions is $2 (P - V + 1 + \alpha)$.
  On the other hand, one index loop can accommodate at most two $\CW$.
  Thus we have $h \geq P - V + 1 + \alpha$.  
  This implies that only the planar diagrams contribute to the effective
  superpotential as the Euler number of the diagram is $\chi = V - P + h$.

  A planar diagram with $h$ index loops has $(h - 1)$ loop momenta.
  Let us consider the $(h-1)$-loop planar diagrams (contributing to the $(h-1)$-loop vacuum amplitude)
  in which all vertices are type I.
  Let us, for a moment, ignore the $M$ term of (\ref{M}).
  The calculation is then the same as that of \cite{DGLVZ} which we briefly describe. 
  Each diagram is a product of the bosonic part obtained by integrating over the momentum $p$ 
  and the fermionic one coming from the $\pi^\alpha$ integrations.
  As we have seen in the last paragraph, we have exactly $2(h-1)$ $\CW$ insertions in the fermion part.
  There are two possibilities for these $\CW$ insertions.
  The one is to keep one of the index loops empty, filling the remaining index loops with two $\CW$.
  This yields $N S^{h-1}$ term, where $S =-\frac{1}{64 \pi^2} \Tr_{U(N)} \CW^\alpha \CW_\alpha$.
  The other is to fill each of two index loops chosen with single $\CW$, 
  which yields $S^{h-2} w^\alpha w_\alpha$ terms where $w^\alpha= \frac{1}{8 \pi} \rm{Tr} \mathcal{W}^\alpha$.
  After calculating the both parts, we perform the Schwinger parameter integrals. 
  Clearly this procedure is universal to every $(h-1)$-loop planar diagram 
  up to the multiplications by the symmetric factor and by the coupling constants.
  Therefore every such diagram is a product of these factors with the following expression
    \bea
    \left(
    \prod_{i=1}^{P} \int d s_i 
    \right)
    e^{-(\sum s_i)m'}
    \frac{1}{4^{h-1}} \{ N h S^{h - 1} + {}_h C_2 2 S^{h - 2} w^\alpha w_\alpha \}
    \equiv 
    \left(
    \prod_{i=1}^{P} \int d s_i 
    \right)
    e^{-(\sum s_i)m'}
    \mathcal{A}_0^{(h-1)},
    \label{amp1}
    \eea
  where we have introduced $\mathcal{A}_0^{(h-1)}$.
  The factor $h$ of the first term comes from the choice of the empty index loop, 
  and ${}_h C_2$ of the second term is the combination of inserting two $\CW$ into different index loops.
  The most important fact is that the dependence on Schwinger parameters of
  the bosonic part is cancelled by that of the fermionic part. 
  This explains that the calculation of the effective superpotential of the gauge theory 
  reduces to that of the matrix model \cite{DGLVZ}.
  
  There are two types of corrections to $\mathcal{A}_0^{(h-1)}$.
  The one is due to the presence of the $M$ terms in the propagators, 
  which we denote by $\mathcal{A}_1^{(h-1)}$.
  The other is due to the type II vertices, 
  which is obtained by replacing one of the type I vertices 
  in $\mathcal{A}_0^{(h-1)}$ by the corresponding type II vertex 
  and by summing over all possibilities.
  We denote this by $\mathcal{A}_2^{(h-1)}$.  We consider them in order.
  
  Let us see the effects of the $M$ term, namely, eq. (\ref{M}).
  It plays a role of inserting two $\CW$ further.
  Thus we will obtain terms which are proportional to $S^h$.
  Note that we cannot insert more than two $\CW$  because, in such case, 
  at least one of the index loops has more than two insertions of $\CW$.
  For the parts contributing to $N S^{h-1}$, which have an empty index loop, 
  we can further insert $\CW^\alpha \CW_\alpha$ from the first two terms in (\ref{M}).
  In the case in which they are inserted in the $a$-th index loop, 
  we obtain
    $\left( \frac{S}{4} \right)^{h-1} i g'_3 
    \left( {\displaystyle \sum_{i_a}} s_{i_a} \right) {\rm Tr} \CW \CW$,
  where $i_a$ labels the propagators which form the $a$-th index loop.
  The absence of factor $N$ is explained by the absence of an empty index loop.
  The factor $h$ is not present as we have so far restricted ourselves to the
  $a$-th index loop. Summing over all index loops, 
  we obtain the first contribution to $\mathcal{A}_1^{(h-1)}$:
    \bea
    \sum_a
    \left(
    \frac{S}{4}
    \right)^{h-1}
    i g'_3 
    \left(
    \sum_{i_a} s_{i_a} 
    \right)
    {\rm Tr} \CW \CW
     =     2 i g'_3
           \left(
           \sum_i s_i 
           \right)
           \left(
           \frac{S}{4}
           \right)^{h-1}
           {\rm Tr} \CW \CW,
           \nonumber
    \eea
  where we have used  that when all index loops are summed, they pass through
  each double line propagator exactly twice.

  Let us note that the parts contributing to the second term of eq. (\ref{amp1}) 
  can receive further insertions of $\CW$ as well.
  They have two index loops with a single $\CW$ insertion, for which we can exploit the last term of $M$.
  An insertion of this term requires that two index loops share a propagator.  
  Let us define the index $A=1, \ldots , {}_h C_2$ as labeling the combinations of such two index loops
  and the index $\tilde{A}$ labeling the cases which have a common propagator in the two index loops chosen.
  Let us further introduce the index $i_{\tilde{A}}$ labeling the common propagator in case $\tilde{A}$.
  With these notations, we obtain the second contribution to $\mathcal{A}_1^{(h-1)}$:
    \bea
    \frac{2 S^{h-2}}{4^{h-1}} i g'_3 
    \left(
    \sum_{i_{\tilde{A}}} s_{i_{\tilde{A}}}
    \right)
    \frac{1}{64 \pi^2} \CW^\alpha_{ab} \CW_{\alpha cd} \CW^\beta_{ba}
    \CW_{\beta dc}
     =     i g'_3
           \left(
           \sum_i s_i 
           \right)
           \left(
           \frac{S}{4}
           \right)^{h-1}
           {\rm Tr} \CW^\alpha \CW_\alpha.
           \nonumber
    \eea 
   
  Putting all these together, we obtain the contributions  from the vertices of type I, 
    \bea
    \left(
    \prod_{i=1}^{P} \int d s_i 
    \right)
    e^{-(\sum s_i)m'}
    (\mathcal{A}_0^{(h-1)} + \mathcal{A}_1^{(h-1)} (s_i) )~~~~~~~~~~~~~~~~~~~~~~~~~~~~~~~~~~
    & &    \nonumber \\
     =     \frac{h}{m'^P} \left(\frac{S}{4}\right)^{h-1} 
           \left(
           N - \frac{16 \pi^2 i P g_3 S}{h m g_2}
           \right)
         + \frac{{}_h C_2}{2 m'^P} \left(\frac{S}{4}\right)^{h-2} w^\alpha w_\alpha.
           \label{Wefftype1}
    \eea
  It is important that the above new term has Schwinger parameter dependence aside from the exponential factor.
  In \cite{DGLVZ}, it was pointed out that the cancellation of
  this dependence represents the reduction of the system to the matrix model.
  The appearance of this new term with Schwinger parameter dependence may spoil this reduction.
  Note also that this new term does not have an overall factor $N$,
  indicating the violation of the well-known relation due to Dijkgraaf-Vafa \cite{DV}.
  
  We now turn to the vertices of type II which contain two $\CW$ insertions.
  The $\ell$-th order vertex in $\Phi$ is
    \bea
    {\rm Tr}
    (2 \CW \CW \Phi^\ell + \CW \Phi \CW \Phi^{\ell - 1} + \ldots + \CW \Phi^{\ell - 1} \CW \Phi).
    \label{type2ell}
    \eea
  where we have omitted the overall factors.
  The first term inserts two $\CW$ into an index loop while 
  the remainder insert them into two different index loops.
  Having done  $2(h-1)$ $\pi^{\alpha}$ integrations, 
  we obtain $2(h-1)$ $\CW$ insertions.
  We can therefore use vertex (\ref{type2ell}) only once in a diagram.
  When this is done, insertion of the $M$ term from the propagator is disallowed.
  
  Let us consider $\mathcal{A}_2^{(h-1)}$ and suppose that one of the type I vertices, 
  $\Tr \Phi^\ell$, is replaced by the above vertex (\ref{type2ell}).
  The first term connects $\ell$ index loops and we can insert $\CW^2$ into $\ell$ different ways.  
  Thus we obtain 
  $ \left( \frac{S}{4} \right)^{h - 1} 2 \ell {\rm Tr} \CW \CW $ as a contribution to $\mathcal{A}_2^{(h-1)}$.
  For the other terms of eq. (\ref{type2ell}), there are in total $\ell(\ell - 1)$ 
  ways of inserting two $\CW$ into different index loops.
  These give
    \bea
    \frac{2 S^{h - 2}}{4^{h-1}} \ell (\ell - 1) 
    \frac{1}{64 \pi^2} \CW^\alpha_{ab} \CW_{\alpha cd} \CW^\beta_{ba} \CW_{\beta dc}
     =     \left(
           \frac{S}{4}
           \right)^{h - 1} 
           \ell (\ell - 1)
           {\rm Tr} \CW \CW.
           \nonumber
    \eea
  Summing the above two contributions, we obtain $ \left( \frac{S}{4} \right)^{h - 1} \ell (\ell + 1) {\rm Tr} \CW \CW $.
  Thus, in any $(h-1)$-loop diagram, changing a vertex from type I to type II 
  is equivalent to considering only $NS^{h-1}$ terms in eq. (\ref{amp1}) and changing the coupling constant by
    \bea
    m g_\ell
    \rightarrow
           \frac{16 \pi^2 i g_{\ell + 1} S}{N h},
           ~~~~~~~{\rm for} ~~\ell \geq 3.
           \label{Wefftype2}
    \eea
  
  Therefore, we obtain a formula for the contribution from the $(h-1)$-loop diagrams with $P$ propagators
  to $W_{eff}$ in (\ref{S2}), 
    \bea
    W_{eff}^{(h-1)}
     =     N \frac{\partial F^{(h-1)}}{\partial S}
         + \frac{\partial^2 F^{(h-1)}}{\partial S^2} w^\alpha w_\alpha
         - \frac{16 \pi^2 i P m g_3}{h m g_2} 
           \left(
           \frac{\partial F^{(h-1)}}{\partial S}
           \right) 
           \frac{S}{m}
         + W_2^{(h-1)},
           \label{Weffh-1}
    \eea
  where $W_2^{(h-1)}$ is defined 
  by replacing, in the first term, one coupling constant according to eq. (\ref{Wefftype2}) 
  and summing over all possibilities.
  We have denoted by $F^{(h-1)}$ the $(h-1)$-loop contribution to the planar free energy of the matrix model.
  
%%%%%%%%%%%%%%%%%%%%%%%%%%%%%%%%%%%%%%%%%%%%%%%%%%%%%%%%%
\section{Example}
\begin{figure}[t]
    \begin{center}
    \includegraphics[scale=0.7]{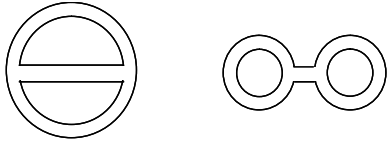}
    \caption{two-loop planar diagrams}
    \label{fig:2-loop}
    \end{center}
    \end{figure}

  As a sample computation, let us take the two-loop contribution
  to the effective superpotential.
  There are two two-loop planar diagrams depicted in Fig.\ref{fig:2-loop}.
  Collecting all possible insertions of $\CW$, we obtain
    \begin{equation}
    W_{eff}^{(2)}
     =   - \frac{(m g_3)^2}{32 (m g_2)^3} N S^2 
         - \frac{(m g_3)^2}{16 (m g_2)^3} S w^\alpha w_\alpha
         + \frac{\pi^2 i (m g_3)^3}{2 (m g_2)^4} \frac{S^3}{m}
         - \frac{\pi^2 i (m g_3) (m g_4)}{3 (m g_2)^3} \frac{S^3}{m}.
           \label{W2loop}
    \end{equation}
  The first two terms are the ones which are present
  in the computation based on \cite{DV,CDSW} with $S_{\CN=1}$. 
  The third one comes from the $M$ term in the propagator and
  the last one from the type II vertices.
  Note that, in the limit $m \rightarrow \infty$ with $m g_k$
  ($k \geq 2$) fixed, we reproduce the result of \cite{DGLVZ}.
  In an arbitrary loop amplitude, the situation is the same: new terms are of order $m^{-1}$ in this limit.

  The overall $U(1)$ part does not decouple from the $SU(N)$ part.
  This can be easily seen by translating $S$ into the glueball superfield 
  $\hat{S} = - \frac{1}{64 \pi^2} \Tr_{SU(N)} \CW^\alpha \CW_\alpha$ 
  and extracting the factor in front of $w^\alpha w_\alpha$.
  By the existence of the last two terms in eq. (\ref{Weffh-1}), it is nonvanishing.
  For example, in the two-loop example, this part in (\ref{W2loop}) reads
    \bea
    \frac{\pi i (m g_3) [2(m g_2)(m g_4) - 3(m g_3)^2 ]}{2 (m g_2)^4} \frac{\hat{S}^2}{m} w^\alpha w_\alpha \neq 0,
    \nonumber
    \eea
  
%%%%%%%%%%%%%%%%%%%%%%%%%%%%%%%%%%%%%%%%%%%%%%%%%%%%%%%%%%%%%%%%%%%%%%%%%%%%%%%%%%%%%%%%%%%%%%%%%%%%%
\section{The chiral ring and the generalized Konishi anomaly}
 An alternative approach to the effective superpotential 
  is to exploit and extend the properties of the $\CN = 1$ chiral ring and
 the generalized Konishi anomaly equations based on reference \cite{Konishi, CDSW}.  
  The anomalous Ward identity of our model for the general transformation
 $\delta \Phi = f(\Phi, \CW)$ is 
    \bea 
         - \left< \frac{1}{64 \pi^2} 
           \left[ 
           \CW^\alpha , 
           \left[ \CW_\alpha , \frac{\partial f}{\partial \Phi_{ij}}
           \right]
           \right]_{ij}
           \right>_{\Phi}   
     =     \left<
           {\rm Tr} f W'(\Phi)
           \right>_{\Phi}
         - \left<
           \frac{i}{4} {\rm Tr} (f \CF'''(\Phi) \CW^\alpha \CW_\alpha)
           \right>_{\Phi},
           \label{WIforg}
    \eea
  where $W''(\Phi) = m \CF'''(\Phi)$.
  In terms of the two generating functions of chiral one-point
  functions
    \bea
    R(z)
    &=&  - \frac{1}{64 \pi^2} \left<
           {\rm Tr} 
           \CW^\alpha \CW_\alpha \frac{1}{z - \Phi}
           \right>_{\Phi},
           \nonumber \\
    T(z)   
    &=&    \left< {\rm Tr} 
           \frac{1}{z - \Phi}
           \right>_{\Phi},
           \nonumber
    \eea
  the anomalous Ward identities (\ref{WIforg}) are
    \bea
    R(z)^2
    &=&    W'(z) R(z) + \frac{1}{4} f(z),
           \nonumber \\
    2 R(z) T(z)
    &=&    W'(z) T(z) + \frac{1}{4} c(z)
         + 16 \pi^2 i \CF'''(z) R(z)
         + \frac{1}{4} \tilde{c}(z)
           \nonumber,
    \eea
  where $f(z)$ and $c(z)$ are polynomials of degree $n-1$ in $z$ and $\tilde{c}(z)$ is a polynomial of degree $n-2$:
    \bea
    f(z)
    &=&  - \frac{1}{16 \pi^2} \Tr 
           \left<
           \frac{(W'(\Phi) - W'(z)) \CW^\alpha \CW_\alpha}{z - \Phi}
           \right>_\Phi,
           \nonumber \\
    c(z)
    &=&    4 \left<
           \frac{W'(\Phi) - W'(z)}{z - \Phi}
           \right>_\Phi,
           \nonumber \\
    \tilde{c}(z)
    &=&  - i \left<
           \frac{(\CF'''(\Phi) - \CF'''(z)) \CW^\alpha \CW_\alpha}{z - \Phi}
           \right>_\Phi.
           \nonumber
    \eea 
  The last term of eq. (\ref{WIforg}) does not contribute to the equation for $R(z)$ 
  because of the chiral ring relation ${\rm Tr} \CW^\alpha \CW_\alpha \CW^\beta \CW_\beta = 0$.
  The equation for $R(z) $ is the same as that of \cite{CDSW}, which is the loop equation of the matrix model.
  On the other hand, the equation for $T(z)$ alters from that of \cite{CDSW}.
    
  The final step of this approach is to express the effective superpotential in terms of $R(z)$ and $T(z)$.
  Taking a variational derivative of (\ref{Weff}) with respect to the coupling $g_k$, we obtain
    \bea
    \frac{\partial W_{eff}}{\partial g_k}
     =     \frac{m}{k!} \int d z z^k T(z)
         + \frac{16 \pi^2 i}{(k-1)!} \int d z z^{k-1} R(z).
           \nonumber
    \eea
  Hence we can determine the effective superpotential up to $g_k$ independent
  terms.

%%%%%%%%%%%%%%%%%%%%%%%%%%%%%%%%%%%%%%%%%%%%%%%%%%%%%%%%%%%%%%%%%%%%%%%%%%%%%%%%%%%%%%%%
\section*{Acknowledgements}
  We thank Kazuhito Fujiwara, Yosuke Imamura, Hiroaki Kanno, Hironobu Kihara, Yasunari Kurita, 
  Kazutoshi Ohta and Makoto Sakaguchi for useful discussions.
  We are grateful to Hiraku Yonemura for his collaboration at an early stage.
  This work is supported in part by the Grant-in-Aid for Scientific Research (18540285) 
  from the Ministry of Education, Science and Culture, Japan.
  Support from the 21 century COE program ``Constitution of wide-angle mathematical basis focused on knots'' 
  is gratefully appreciated.
  The preliminary version of this work was presented in YITP workshop 
  ``Fundamental Problems and Applications of Quantum Field Theory'', YITP-W-06-16
  in Yukawa Institute for Theoretical Physics, Kyoto University (December 14-16 2006).
  We wish to acknowledge the participants for stimulating discussions.

%%%%%%%%%%%%%%%%%%%%%%%%%%%%%%%%%%%%%%%%%%%%%%%%%%%%%%%%%%%%%%%%%
%%% references
%%%%%%%%%%%%%%%%%%%%%%%%%%%%%%%%%%%%%%%%%%%%%%%%%%%%%%%%%%%%%%%%%

\end{document}